\documentclass[twoside]{dis09}
\usepackage[latin1]{inputenc}
\usepackage[dvips]{graphicx,epsfig,color}
\usepackage{wrapfig,rotating}
\usepackage{amssymb,amsmath,array}

\input babarsym.tex
\def\CP{$ C \! P$ }

\def\ra{\rightarrow}
\def\babar{\mbox{\slshape B\kern-0.1em{\smaller A}\kern-0.1em B\kern-0.1em{\smaller A\kern-0.2em R}}~} 
\def\babarn{\mbox{\slshape B\kern-0.1em{\smaller A}\kern-0.1em B\kern-0.1em{\smaller A\kern-0.2em R}}} 

\begin{document}

\title{Results on Searches for New Physics at B Factories}

\author{Gerald Eigen, representing the BABAR collaboration
%
\thanks{I would like to thank the \babar collaboration, in particular K. Schubert for their help.  This work has been supported by the Norwegian Research Council.
}
%
\vspace{.3cm}\\
%
University of Bergen - Dept of Physics \\
Allegaten 55, Bergen, Norway
}

\maketitle

\begin{abstract}

We summarize recent results on $B^+ \ra \tau^+ \nu_\tau$ setting constraints on the charged Higgs mass,  discuss the \CP puzzle in $B \ra K \pi $ decays and present searches for a light neutral Higgs in radiative $\Upsilon(2S)$ and $\Upsilon(3S)$ decays. 
\end{abstract}

\section{Introduction}

Rare decays are processes with branching fractions of ${\cal O}(10^{-4})$ or smaller. Typically, they arise if amplitudes of higher-order processes (penguin loops, box diagrams) become dominant because tree amplitudes are suppressed in the Standard Model (SM). Additional suppression comes from
small CKM couplings and helicity conservation. Contributions of New Physics (NP) processes may become significant modifying the prediction with respect to those in the SM. Thus, rare decays provide an interesting hunting ground for NP searches that are complementary to direct searches at the LHC.  

\section{Measurement of ${\cal B}(B^+ \ra \tau^+ \nu_\tau)$}

In the SM, $B^+ \ra \tau^+ \nu_\tau$\footnote{charge conjugation is implied unless otherwise stated} proceeds via $W$ annihilation which is helicity-suppressed. The branching fraction involves
the $B$ decay constant ($ f_B$), the CKM matrix element $(|V_{ub}|)$, and the $\tau$ mass. For a recent lattice result of $\rm f_B = 195\pm 11~MeV$ \cite{fb} and $|V_{ub}|=(3.93\pm0.36)\times 10^{-3}$\cite{pdg} the SM prediction yields ${\cal B}(B^+ \ra \tau^+ \nu_\tau) =(1.04 \pm 0.19 \pm 0.12) \times 10^{-4}$, where the uncertainties originate from the errors in $|V_{ub}|$ and $f_B$, respectively. 
In extensions of the SM, an additional contribution may arise from a charged Higgs boson modifying the branching fraction by an additional factor  \cite{hou}
\begin{equation}
r_H=\bigl ( 1- \frac{m_B^2}{m_H^2}\frac{ \tan^2 \beta}{1+\epsilon_0 \tan \beta }\bigr ) ^2,
\end{equation}
where $m_H$ ($m_B$) is the mass of the charged Higgs boson ($B^+$ meson), $\epsilon_0\simeq 0.01$ is an effective coupling~\cite{rh}  and $\tan \beta $ is the ratio of vacuum expectation values of the two Higgs doublets. 

Both \babar and Belle looked for $B^+ \ra \tau^+ \nu_\tau$ analyzing 467 Million and 657 million $B \bar B $ events, respectively. One $B$ meson, called a "tag", is fully reconstructed in semileptonic $B \ra D^0 \ell \nu X$ or hadronic decays. In the recoil one looks for decays $\tau^+ \ra e^+  \nu_e \bar \nu_\tau$, $\tau^+ \ra  \mu^+  \nu_\mu \bar  \nu_\tau$, $\tau^+ \ra \pi^+  \bar \nu_\tau$ and  $\tau^+ \ra \rho^+  \bar \nu_\tau$, thus selecting events with an isolated $e^+, \mu^+, \pi^+$ or $\pi^+ \pi^0$ in the recoil.  For signal candidates the extra neutral energy in the event, $\rm E_{extra}$,  is examined. This is the energy of all photons in the electromagnetic calorimeter that do not belong to the signal nor the reconstructed tag. For correctly reconstructed tags $E_{extra}$ represents the summed noise in the calorimeter. Double semileptonic tags are used to cross check the simulation.

\begin{wrapfigure}{r}{0.5\columnwidth}
\vskip -0.6 cm
\centerline{\includegraphics[width=0.47 \columnwidth]{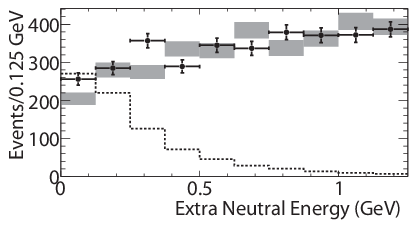}} \vskip -0.5 cm
\caption{Distribution of extra neutral energy for semileptonic tags from \babar showing data (points), expected background (shaded histogram) and expected signal (dotted line).}\label{Fig:eextra} \vskip 0.3 cm
\centerline{\includegraphics[width=0.55\columnwidth]{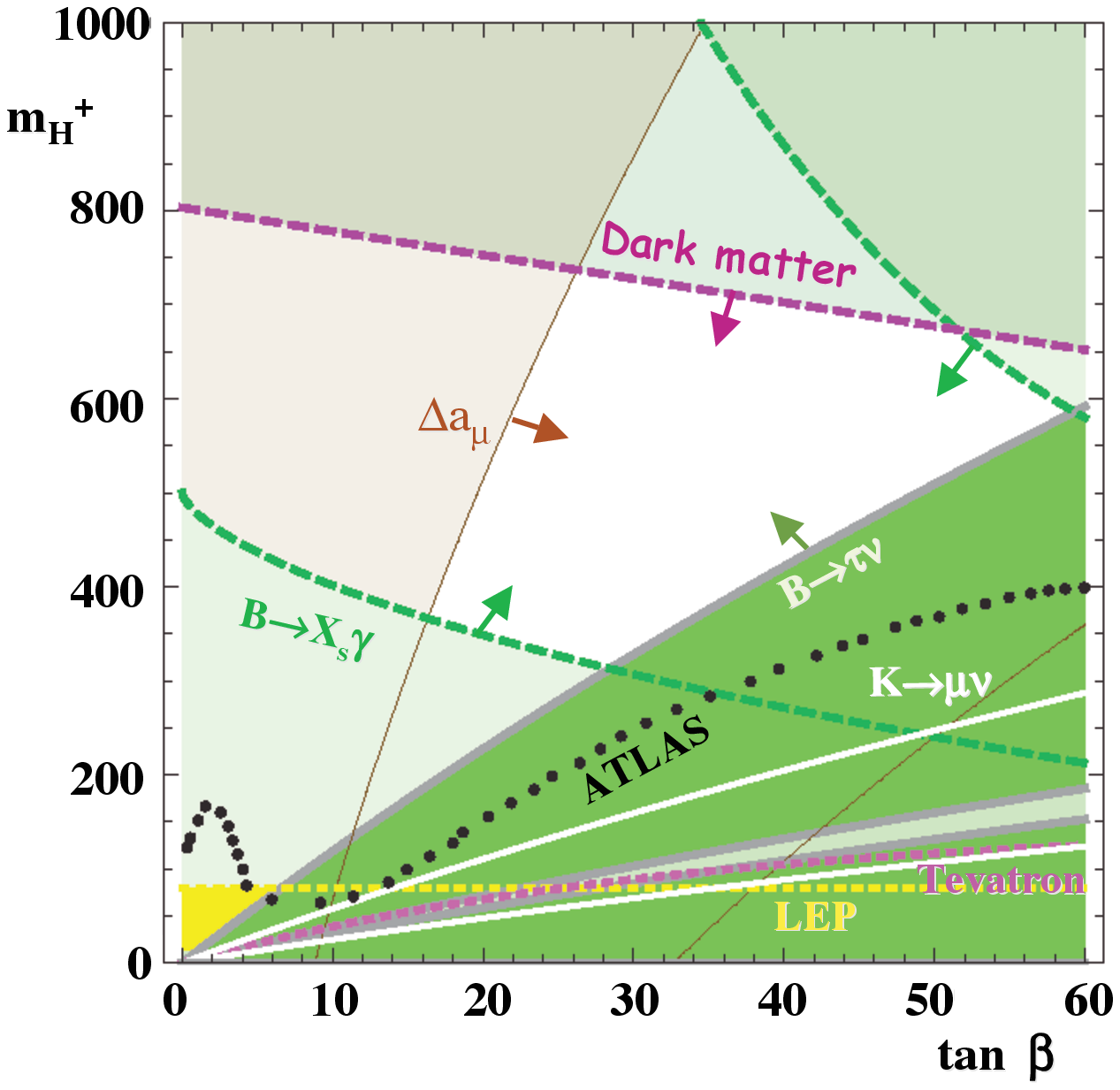}}\vskip -0.2cm
\caption{Exclusion regions  at $95\% ~C.L.$ in the $m_H - \tan \beta$ plane from measurements of ${\cal B}(B^+ \ra \tau^+ \nu_\tau)$ (thick solid lines),  ${\cal B}(K^+ \ra \mu^+ \nu_\mu)$ (light solid lines), ${\cal B}(B \ra X_s \gamma)$ (lower an upper right thick dashed lines), dark matter searches (upper thick dashed line), $\Delta a_\mu$ (left and right thin solid lines), charged Higgs searches at LEP (thin light dotted line), charged Higgs searches at the Tevatron (thin dark dotted line) and the $5 \sigma$ discovery curve in ATLAS for $30~fb^{-1}$ (thick dotted line).}\label{Fig:mh-tb}
\end{wrapfigure}

The $E_{extra}$  distribution measured in \babar is shown in Figure~\ref{Fig:eextra}. We extract signal in the region $\rm E_{extra} < 0.2~ GeV$ after extrapolating background from a sideband ($\rm E_{extra} >0.6~GeV$). We see an excess of $89 \pm 44$ events. The total selection efficiency is $\epsilon =(1.18\pm 0.1)\times 10^{-3}$.  Including the yields of a previous analysis using hadronic tags we measure a $3.2 \sigma$ significant branching fraction of 
${\cal B}( B^+ \ra \tau^+ \nu_\tau)= (1.8 \pm 0.6_{stat} \pm 0.1_{sys}) \times 10^{-4}$~\cite{babar1}.

Belle sees an excess of $154{^{+36}_{-35}}{^{+20}_{-22}}$ events measuring a branching fraction of
${\cal B}(B^+ \ra \tau^+ \nu_\tau) = (1.65^{+0.38+0.35}_{-0.37-0.37}) \times 10^{-4}$
($3.8 \sigma$ significance)~\cite{belle1}. This is in good agreement with ${\cal B}(B \ra \tau \nu) = (1.79^{+0.56+0.46}_{-0.49-0.51}) \times 10^{-4}$ measured in hadronic tags using 449 million $B\bar B$ events~\cite{belle2}.

Accounting for correlated systematic uncertainties the \babarn /Belle average  is
${\cal B}(B^+ \ra \tau^+ \nu_\tau) = (1.73\pm 0.37) \times 10^{-4}$ ( $4.6 \sigma$ significance). Division by the SM branching fraction yields $r_H =1.67 \pm 0.36 \pm 0.36$, where the first error gives the total experimental uncertainty and the second error accounts for uncertainties in $f_B$ and $|V_{ub}|$. 

From the measurement of $r_H$ we determine $\rm 95\%$~confidence level (C.L.)  contours in  $m_H - \tan \beta$ plane that are shown in Figure~\ref{Fig:mh-tb} for the Minimal Supersymmetric Standard Model (MSSM) with minimum flavor violation. In addition, we show $\rm 95\%$ C.L. contours for ${\cal B}(K^+ \ra \mu^+ \nu_\mu)$ ~\cite{pdg}, ${\cal B}(B \ra X_s \gamma)$~\cite{pdg}, $\Delta a_\mu$ (difference of measured anomalous magnetic moment of the muon and the SM prediction, $1.2\times 10^{-9} < \Delta a_\mu < 4.6 \times 10^{-9}$)~\cite{amu}, dark matter searches ($0.079 < \Omega_{CDM}  h^2< 0.119$) \cite{dm}, and Higgs searches at LEP and at the Tevatron \cite{pdg} that are calculated for a heavy-squark scenario ($\rm M_{\tilde q} =1. 5~TeV, A_U=-1~TeV, \mu=0.5~TeV$ and $\rm M_{\tilde \ell}=0.4~TeV$)~\cite{isidori}. Note that ${\cal B}(B^+ \ra \tau^+ \nu_\tau)$ and ${\cal B}(K^+ \ra \mu^+ \nu_\mu)$ already exclude a large part of the $m_H - \tan \beta$ plane. The other constraints come from dark mater searches, $\Delta a_\mu$ and ${\cal B}(B \ra X_S \gamma)$. The allowed region is the white area located in the center of the plot. The dotted curve shows the $5 \sigma$ discovery curve expected in ATLAS for $30 fb^{-1}$~\cite{atlas}. For other SUSY parameters, the allowed area typically shrinks~\cite{isidori}.

\section{The \CP Puzzle in $B \ra K \pi$ Decays}

\begin{figure}
\centerline{\includegraphics[width=0.99 \columnwidth]{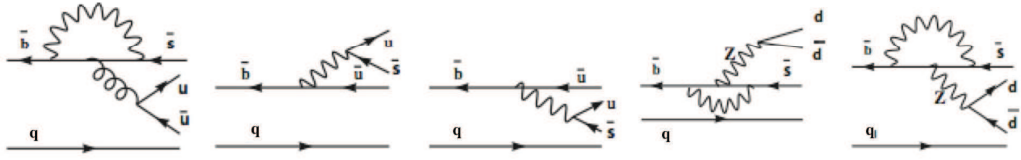}}
\caption{Lowest-order diagrams for $B \ra K \pi$ decays, (from left to right) gluonic penguin, tree, color-suppressed tree, color-allowed EW penguin and color-suppressed EW penguin.} \label{Fig:kpi}
\end{figure}

The decays $B \ra K \pi$ are dominated by gluonic penguin amplitudes with a t-quark in the loop ($P'_t \sim {\cal O}(1)$). Tree amplitudes  ($T'$) are suppressed by ${\cal O} (\lambda)$, while color-suppressed tree ($C'$) and gluonic penguin amplitudes with a u-quark in the loop ($P'_u$) are suppressed by ${\cal O} (\lambda^2)$, where $\lambda = 0.22$. In addition, electroweak (EW) penguin ($P'_{EW}$) and color-suppressed EW penguin amplitudes ($P'^C_{EW}$) contribute at  ${\cal O} (\lambda)$ and ${\cal O} (\lambda^2)$, respectively~\cite{gronau1}.

\begin{wrapfigure}{r}{0.5\columnwidth}
\vskip - 0.3 cm
\centerline{\includegraphics[width=0.40\columnwidth]{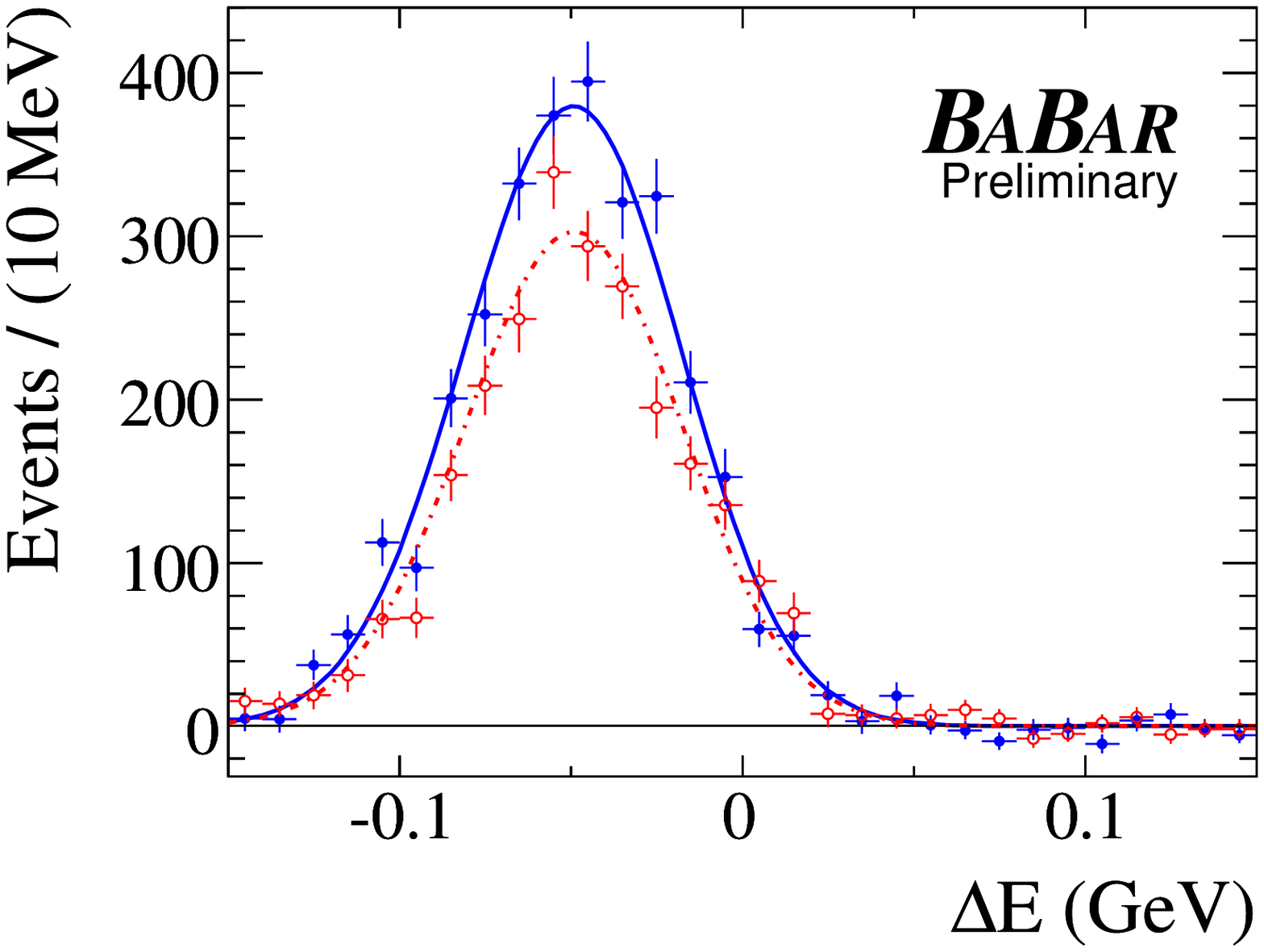}} \vskip -0.3 cm
\caption{$\Delta E$ distributions for $B^0 \ra K^+ \pi^-$ (solid) and  $\bar B^0 \ra K^- \pi^+$ (dashed) decays.}\label{Fig:cpkpi}
\centerline{\includegraphics[width=0.48\columnwidth]{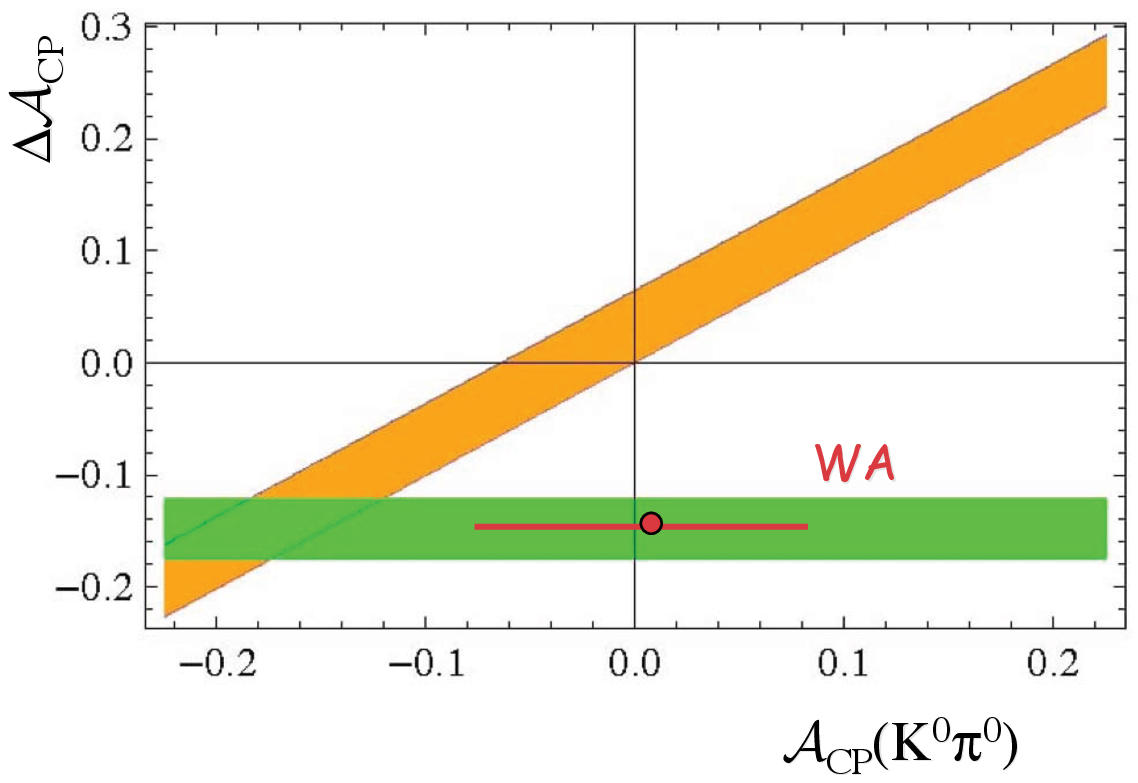}} \vskip -0.2cm
\caption{SM sum rule relating $\Delta {\cal A}_{K\pi}$ to $ {\cal A}(K^0 \pi^0)$ (diagonal band), WAs for $\Delta {\cal A}_{K\pi}  $ (horizontal band) and ${\cal A}(K^0 \pi^0)$ (point).}\label{Fig:sr}\vskip -0.3 cm
\end{wrapfigure}

In $B^+ \ra K^+ \pi^0$ and  $B^0 \ra K^0 \pi^0$ decays all processes shown in Figure~\ref{Fig:kpi} contribute, while in $B^0 \ra K^+ \pi^-$ and $B^+ \ra K^0 \pi^+$ decays the amplitudes
$P'_{EW}$ and $C'$ are absent. Since $P'_{EW}$ is at ${\cal O}(\lambda)$, branching fractions might  differ substantially due to interference. The measured ratios of branching fractions corrected for isospin and different $B^+$ ($\tau_+ )$ and $B^0$ ($\tau_0$) lifetimes are close to one: $\frac {{\cal B}(B^0 \ra K^+ \pi^-)}{2 \times {\cal B}(B^+ \ra K^+ \pi^0)} \frac{\tau_+}{\tau_0}=0.81\pm 0.05 $ and $\frac {2 \times {\cal B}(B^0 \ra K^0 \pi^0)}{{\cal B}(B^+ \ra K^0 \pi^+)} \frac{\tau_+}{\tau_0}=0.91\pm 0.07$. The largest deviation is less than $20\%$, indicating that contributions from $P'_{EW}$ and $C'$ are small. 

Direct \CP violation is another interesting observable.
\babar updated direct \CP violation measurements for $B^0 \ra K^+ \pi^-$ with the full data set. As shown in Figure~\ref{Fig:cpkpi} we see a large \CP asymmetry of  ${\cal A}(K^+\pi^-) = -0.107\pm 0.016^{+0.006}_{-0.004}$ ~\cite{babar2}. This  increases the world average (WA) to ${\cal A}(K^+\pi^-) = -0.098^{+0.012}_{-0.011}$ ($8.1 \sigma $ significant). For $ B^+ \ra K^+ \pi^0 $ the WA is $ {\cal A}(K^+\pi^0) = 0.05\pm 0.025 $. Though consistent with zero at the $2.0 \sigma$ level the difference in \CP asymmetries between $B^0$ and $B^+$ decays is increased to 
$\Delta {\cal A}_{K\pi} = {\cal A}(K^+ \pi^-) - {\cal A}(K^+ \pi^0) =-0.148 \pm 0.028$. Such a large effect ($5.3 \sigma$) is unexpected. The discrepancy $\Delta {\cal A}_{K\pi} $ was attributed to either with a large $C'$~\cite{cpd} or large $P"_{EW}$ contribution~\cite{ewp}. An enhanced $C'$ is due to a strong interaction effect, while an enhanced $P'_{EW}$ may hint at NP in loop processes containing a charged Higgs boson or supersymmetric particles.

The measured value of ${\cal A}(K^0 \pi^+)= 0.009\pm 0.025$ is consistent with zero.  For $B^0 \ra K^0 \pi^0$ both \babar and Belle updated results. Observing $556\pm 32 K^0_S \pi^0$ events in the full data set \babar measures  ${\cal A}(K^0 \pi^0)= -0.13\pm 0.13\pm 0.03$. Belle measures ${\cal A}(K^0 \pi^0)= 0.14\pm 0.13\pm 0.06$ by combining $657\pm37 K^0_S \pi^0$ and  $ 285\pm 77 K^0_L \pi^0$ events. The resulting WA of  ${\cal A}(K^0 \pi^0)= 0.01\pm 0.10$ is in good agreement with ${\cal A}(K^0 \pi^+)$, though errors are large.

In the SM a sum rule connects all four \CP asymmetries ~\cite{soni, gronau} by 
\vskip -0.2 cm
\begin{equation}
{\cal A}(K^+\pi^-) +{\cal A}(K^0\pi^+) \frac{{\cal B}(K^0\pi^-)}{{\cal B}(K^+\pi^-)}\frac{\tau_+}{\tau_0}={\cal A}(K^+\pi^0) \frac{2{\cal B}(K^+\pi^0)}{{\cal B}(K^+\pi^-)}\frac{\tau_+}{\tau_0} +{\cal A}(K^0\pi^0) \frac{2{\cal B}(K^0\pi^0)}{{\cal B}(K^+\pi^-)}. 
\end{equation}
\vskip -0.2 cm
\noindent
Figure~\ref{Fig:sr} depicts the relation between $\Delta {\cal A}_{K \pi}$ and ${\cal A}(K^0\pi^0)$ graphically. If no NP is present, the $K^0 \pi^0$  \CP asymmetry is determined by the overlap region
of the horizontal and diagonal bands yielding ${\cal A}(K^0\pi^0)=-0.151±0.043$. The present WA is consistent with this value at the $1.4 \sigma$ level.

A recent study of the $K \pi$ system shows that all measurements are consistent with the SM at the $\rm20\% ~C.L.$~\cite{beak}. NP may contribute in gluonic penguin or electroweak penguin loops. For the first scenario one still expects ${\cal A}(K^0\pi^0)=-0.15$, while for the second scenario ${\cal A}(K^0\pi^0)=-0.03$. The precision of present data is not sufficient to distinguish between both scenarios. A factor of ten more data is required to settle the issue.  Since LHCb cannot measure the $K^0 \pi^0$ mode,  a Super B-factory is needed to produce precise measurements  \cite{superb}.

\section{Search for Neutral Light Higgs in $\Upsilon$ Decays}

\begin{wrapfigure}{r}{0.5\columnwidth}
\centerline{\includegraphics[width=0.49\columnwidth]{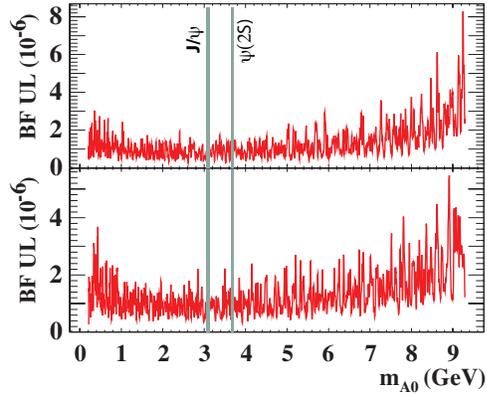}}
\caption{Branching fraction upper limits at $\rm 90\%~C.L.$ for $\Upsilon(2S) \ra  \gamma A^0, A^0 \ra \mu^+ \mu^-$ (top) and  $\Upsilon(3S) \ra  \gamma A^0, A^0 \ra \mu^+ \mu^-$ (bottom).}\label{Fig:a0mm}
\end{wrapfigure}

In the simplest extension of the Minimal Supersymmetric Standard Model (MSSM),  called NMSSM, a \CP-odd Higgs singlet $A_S$ is introduced that mixes with the MSSM \CP-odd state $A_{MSSM}$~\cite{gunion}. The physical state is $A^0 = A_S \sin \theta_A + A_{MSSM} \cos \theta_A$, where $\theta_A$ is the mixing angle. Searches at LEP imposed a lower bound of $| \cos \theta_A | \geq 0.04$ for $\tan \beta=10$.  For decays  $A^0 \ra \tau^+ \tau^-$   the coupling is proportional to $\cos \theta_A \tan \beta$.
Since the $A^0$ may be produced in radiative $\Upsilon$ decays with branching fractions predicted as large as $\rm few \times10^{-4}$, we searched  in \babar for $\Upsilon(3S) \ra \gamma A^0$, $A^0 \ra (\mu^+ \mu^-, \tau^+ \tau^-$, or~invisible) using $(121.8\pm 1.2) \times 10^6~ \Upsilon(3S)$ decays and $\Upsilon(2S) \ra \gamma A^0$, $A^0 \ra \mu^+ \mu^-$ using $(98.6 \pm  0.9) \times 10^6~ \Upsilon(2S)$ decays ~\cite{babar4}.

\begin{wrapfigure}{r}{0.5\columnwidth}
\vskip -0.2 cm
\centerline{\includegraphics[width=0.50\columnwidth]{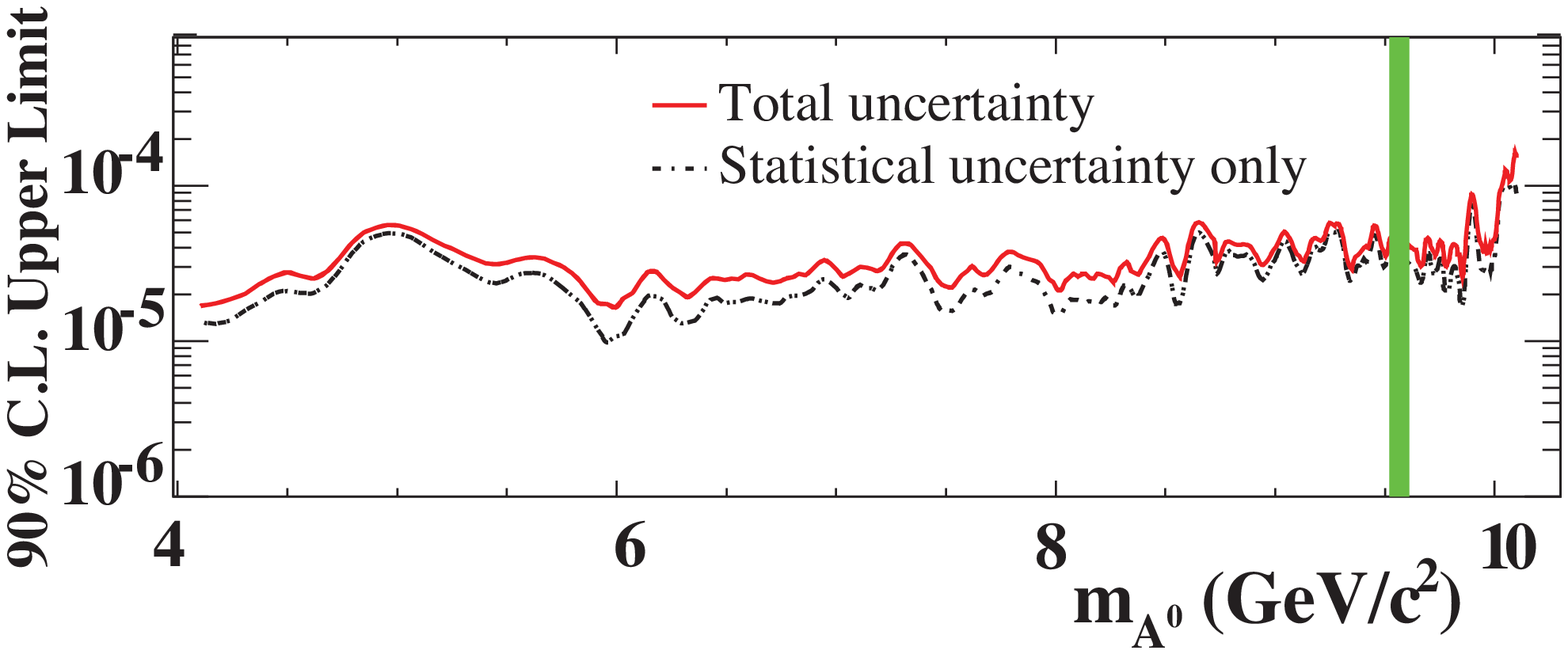}} \vskip -0.2 cm
\caption{Branching fraction upper limits at $\rm 90\%~C.L.$ for $\Upsilon(3S) \ra  \gamma A^0, A^0 \ra \tau^+ \tau^-$ }\label{Fig:a0tt}
\vskip 0.3 cm
\centerline{\includegraphics[width=0.42\columnwidth]{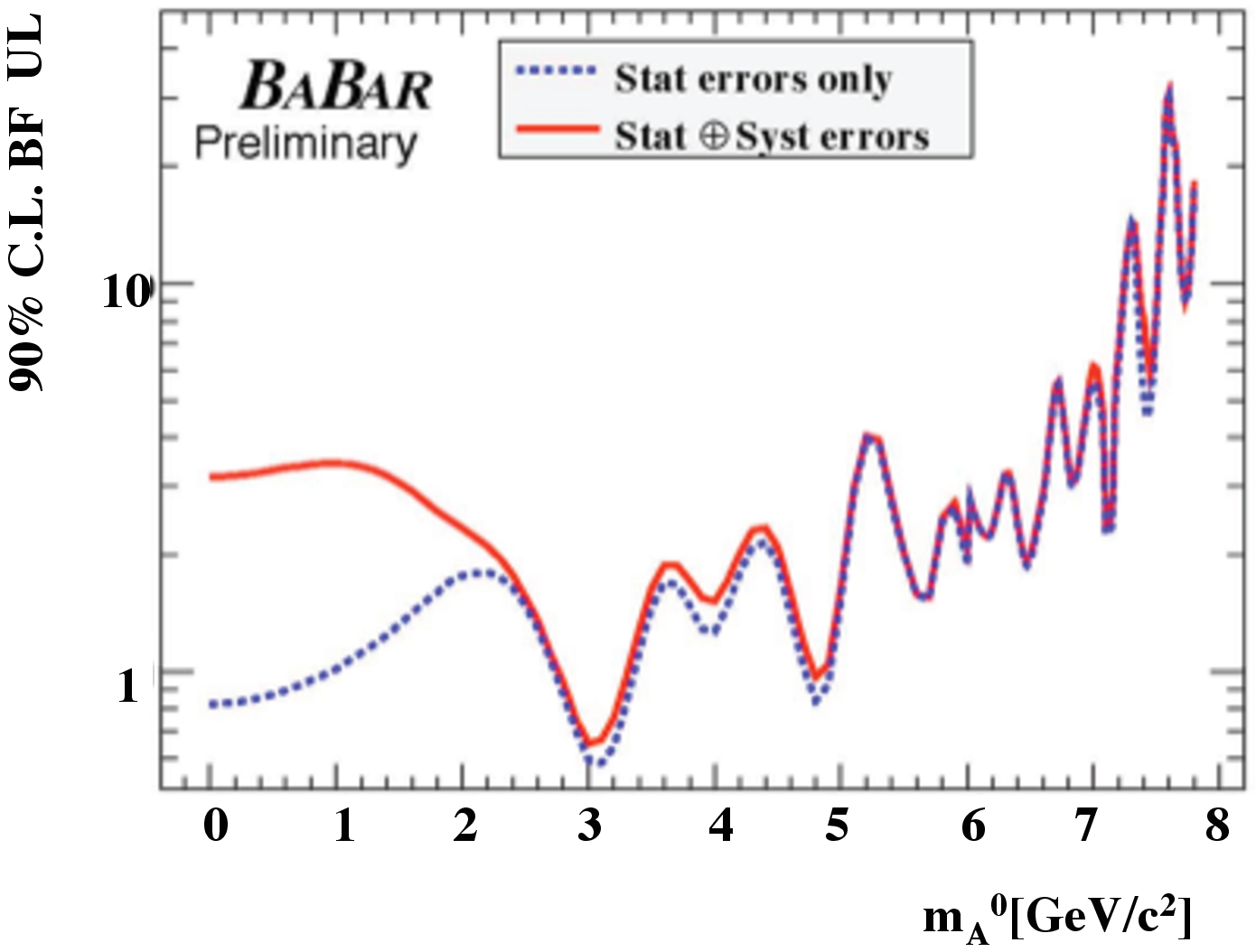}}\vskip -0.2 cm
\caption{Branching fraction upper limits at $\rm 90\%~C.L.$ for $\Upsilon(3S) \ra  \gamma A^0, A^0 \ra \rm  invisible$.}\label{Fig:a0inv}
\end{wrapfigure}

We select events with a photon recoiling against two charged tracks with zero net charge or large missing energy. The latter topology covers the $A^0$ decay into a pair of lightest SUSY particles that escape detection. For the $\mu^+ \mu^-$ decay, we require two identified muons for $\rm m_{\mu\mu}< 1.05~GeV$ to remove $\rho^0$ background. 

We perform a kinematic fit at the $\Upsilon(2S), ~\Upsilon(3S)$ and calculate the reduced mass $m_r =\sqrt{m_{\mu \mu} -4 m_\mu^2}$, for which continuum background from  $e^+ e^- \ra \gamma \mu^+ \mu^-$ is smooth near threshold. Fitting the $\mu^+ \mu^-$ mass spectrum in $300~$MeV bins we see no signal in the entire mass region from $2 m_\mu$  to $\rm 9.3~GeV$. Thus, we set branching fraction upper limits at $\rm 90\%~C.L.$ shown in Figure~\ref{Fig:a0mm}. The limits vary from $0.27 (0.26) \times 10^{-6}$ to $5.5 (8.3) \times 10^{-6}$ for $\Upsilon(3S) (\Upsilon(2S)) $ decays.

In the search for $A^0 \ra \tau^+ \tau^-$, we only select $e^+ e^-$, $\mu^+ \mu^-$ and $e^\pm \mu^\mp$ decays in which both leptons are identified. We look for an excess in a narrow region in the $E_\gamma$ spectrum. We observe no significant signal and set branching fraction upper limits at $\rm 90\% ~C.L.$ shown in Figure~\ref{Fig:a0tt}. The upper limits vary from $15 \times 10^{-6} $ to $ 160 \times 10^{-6}$ being about an order of magnitude larger than those in the $\mu^+ \mu^-$ mode. In the search for $A^0 \ra \rm invisible$, event selection is optimized separately for photon energies $\rm 2.2 < E^*_\gamma < 3.7~GeV$ and $\rm 3.2 < E^*_\gamma < 5.5~GeV$. In both regions an unbinned extended maximum likelihood fit is performed to the distribution of missing mass squared in steps of $\rm \Delta m_{A^0} = 0.1~GeV$. We see no significant signal in the entire mass region  $\rm 0 < m_{A^0} < 7.8~GeV$  and set branching fraction upper limits at $\rm 90\%~ C.L.$ shown in Figure~\ref{Fig:a0inv}, which range from $0.7\times 10^{-6} $ at 3~GeV to $31 \times 10^{-6} $ at 7.6~GeV.


\section{Conclusion}

The large samples of \babar and Belle made it possible to study rare decays. Both experiments found evidence for  $B^\pm \ra \tau^\pm \nu_\tau$. The measured branching fraction is higher than the SM prediction placing stringent constraints on the mass of the charged Higgs boson. In the $B \ra K \pi$ system, \CP asymmetries between $K^+ \pi^-$ and $K^+ \pi^0$  modes differ unexpectedly by $5.3\sigma$. The key issue is a precision measurement of ${\cal A}_{CP}(K^0 \pi^0)$, since this will tell us if the SM holds up or NP is needed.
Using radiative $\Upsilon(3S)$ decays we see no signal for a light neutral Higgs boson in the entire mass region. For a considerable improvement of these measurements a Super B-factory is needed.


\begin{thebibliography}{99}
\bibitem{fb} C. Bernard {\it et~al.}, PoS LATTICE2008, 278 (2008). 
\bibitem{pdg} C. Amsler  {\it et~al.}, Phys.Lett.{\bf B667}, 1 (2008).
\bibitem{hou} W.S. Hou, Phys.Rev.{\bf D48}, 2342 (1993).
\bibitem{rh} A.G.Akeroyd, S. Recksiegel J.Phys.{\bf G29}, 2311 (2003); G. Isidori and P. Paradisi, Phys.Lett.{\bf B639} 499, (2006). 
\bibitem{babar1} \babar Coll. (B. Aubert {\it et~al.}), hep-ex/0809.4027;
Phys. Rev.{\bf  D77}, 011107 (2008).
\bibitem{belle1} Belle Coll. (I. Adachi {\it et~al.}), hep-ex/0809.3834 (2008).
\bibitem{belle2} Belle Coll. (K. Ikado {\it et~al.}), Phys.Rev.Lett. {\bf 97}, 251802 (2006).
\bibitem{bsg} M.Misiak {\it et~al.},  Phys.Rev.Lett. {\bf 98}, 022002 (2007);  G. Degrassi {\it et~al.}, JHEP 0012, 009 (2000). 
\bibitem{amu} G.W. Bennett {\it et~al.}, Phys.Rev.{\bf D73}, 072002 (2006); F.Jegerlehner and A. Nyffeler, Phys.Rept.{\bf 477}, 1 (2009). 
\bibitem{dm} WMAP  Coll. (J. Dunkley {\it et al.}), Astrophys.J.Suppl. {\bf 180}, 306 (2009).
\bibitem{isidori}G. Isidori {\it et~al.}), Phys.Rev.{\bf D75},115019, (2007).
\bibitem{atlas} ATLAS Coll.  (G. Aad {\it et~al.}), hep-ex/0901.0512.
\bibitem{gronau1} M. Gronau,  {\it et~al.}, Phys.Rev.{\bf D50}, 4529,1994; Phys.Rev. {\bf D52}, 6350 (1995); M. Gronau, J. Rosner, Phys. Lett. {\bf B572}, 43 (2003).
\bibitem{babar2} \babar Coll. (B. Aubert {\it et~al.}), hep-ex/0807.4226 (2008). 
\bibitem{babar3} \babar Coll. (B. Aubert {\it et~al.}), Phys.Rev.{\bf D79}, 052003 (2009); hep-ex/0807.4226 (2008).
\bibitem{belle3} Belle Coll. (I. Adachi  {\it et~al.}), hep-ex/0809.4366 (2008).
\bibitem{cpd} C.W.  Chiang   {\it et~al.}, Phys.Rev.{\bf D70}, 034020 (2004);
H.N. Li {\it et~al.}, Phys.Rev.{\bf D72}, 114005 (2005).
\bibitem{ewp} S. Mishima and T. Yoshikawa Phys.Rev.{\bf D70} 094024 (2004);
A.J. Buras  {\it et~al.}, Nucl.Phys.{\bf B697}, 133  (2004); A.J. Buras  {\it et~al.}, Eur.Phys.J.{\bf C45}, 453 (2006); S. Beak and D. London, Phys.Lett.{\bf B653}, 249 (2007);
W.S. Hou {\it et~al.} (2007); Th. Feldmann {\it et~al.}, JHEP{\bf 0808}, 066 (2008).
\bibitem{soni} D. Atwood and  A. Soni, Phys.Rev.{\bf D58}, 036005 (1998).
\bibitem{gronau} M. Gronau, Phys.Lett.{\bf B627}, 82 (2005).
\bibitem{beak} S. Baek {\it et~al.}, Phys.Lett.{\bf B678}, (2009).
\bibitem{superb} M. Bona {\it et~al.} hep-ex/0709.0451 (2007); S. Hashimoto {\it et~al.}, KEK-2004-4 (2004).
\bibitem{gunion} Gunion  {\it et~al.}, Phys.Rev.{\bf D76}, 051105 (2007), R.Dermisek and J. Gunion, Phys.Rev.{\bf D79} 055014 (2009). 
\bibitem{babar4}  \babar Coll.  (B. Aubert {\it et~al.}),  hep-ex/0808.0017 (2008); 0902.2176; 0906.2219 (2009).

\end{thebibliography}
\end{document}